\documentclass[twocolumn,amsmath,
  amssymb,english,aps,prb,floatfix,showpacs]{revtex4}
\usepackage[english]{babel}
\usepackage{times}
\usepackage{graphicx}
\def \mb{\begin{displaymath}} 
\def \me{\end{displaymath}} 
\def \eb{\begin{equation}} 
\def \ee{\end{equation}}   
\def\expect#1{\mathinner{\langle{#1}\rangle}}

{\catcode`\|=\active 
  \gdef\expect#1{\left<\mathcode`\|"8000\let|\bravert {#1}\right>}}
\def\bravert{\egroup\,\vrule\,\bgroup}

\def\dag{\dagger}

\begin{document}

\title{Conductance of a molecule with a center of mass motion}
\author{J. Mravlje$^{1}$, A. Ram\v{s}ak$^{2,1}$, and T. Rejec$^{1,2,3}$}
\affiliation{$^{1}$Jo\v{z}ef Stefan Institute, Ljubljana, Slovenia}
\affiliation{$^{2}$Faculty of Mathematics and Physics, University of Ljubljana,
Slovenia}
\affiliation{$^{3}$Department of Physics, Ben-Gurion University, Beer-Sheva,
Israel}

\begin{abstract}
We calculate the zero temperature conductance and characteristic
correlation functions of a molecule with a center of mass (CM)
motion which modulates couplings to the leads. In the first model studied,
the CM vibrational mode is 
simultaneously coupled to the electron density on the molecule. The
conductance is suppressed in regimes corresponding
to non-integer occupancy of the molecule.  In the second model, where
the CM mode  is not directly coupled to the electron density, the
suppression of conductance is related to the dynamic breaking of the inversion symmetry.
\end{abstract}


\maketitle
\section{INTRODUCTION}
Interest in systems with electron-phonon interaction
has recently reemerged, as a consequence of advances in 
experimental techniques enabling the investigation of the subtle interplay of
electronic and vibrational degrees of freedom by means of measuring
the conductance of the
molecules \cite{madhavan98, park00,park02,
  liang02,zhitenev02,yu04,yu04_2,pasupathy05,zhao05,zhao05_2,
  wahl05,yu05,natelson06}. 
Consequently,
a considerable amount of work was done in pursuit of understanding the equilibrium
 \cite{cornaglia04, cornaglia05a, cornaglia05b,mravlje05} and
non-equilibrium \cite{mitra04, paaske05, takahiro05} properties of 
these systems. The analysis of
low-temperature properties of such systems is challenging as strong
repulsive interaction among electrons
confined to molecular orbitals leads to surprising results such as the
interplay of the Kondo physics and
molecular vibrations, which are intractable to conventional
perturbational approaches. Investigation of the 
molecules which, when attached to the leads,
may posses both internal and center-of-mass vibrational
modes, makes the calculation of various properties even more
involved. 

\begin{figure}
\begin{center}\includegraphics[%
  width=70mm,
  keepaspectratio]{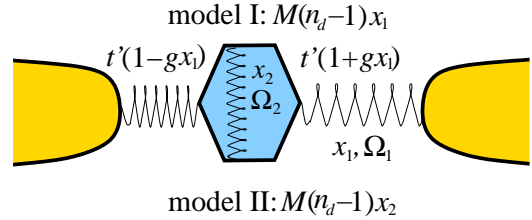}\end{center}

\caption{\label{Fig1} (Color online) Schematic  plot of the
  model devices. In model I the CM vibrational mode which modulates the
  tunneling amplitude  $t'$ is coupled also to the
  local charge density $n$. In model II the CM displacement modulates
  only the tunneling amplitude whereas another (breathing) vibrational mode is
  coupled to the local charge density.} 
\end{figure}

Here we consider a molecule undergoing a linear transport measurement -- a 
system which consists of a molecule and
attached leads. The shape of the molecule and its position with
respect to the leads oscillate (the setup is sketched in 
Fig.\ref{Fig1}). Hence the orbital level of the molecule -- for
transport through 
molecules it is sufficient to consider a
single orbital due to wide inter-level spacings \cite{yu04_2} -- and the
tunneling amplitude are both modulated.
Here we treat two possibilities: (i) in case I (proposed in
Ref.~\onlinecite{alhassanieh05}) the vibrational mode
that modulates the tunneling amplitude 
is also coupled to the electron density at the molecule; (ii) 
in case II (treated also in Ref.~\onlinecite
{balseiro06})   the
molecule possesses in  addition to the CM mode a breathing
mode which modulates the energy level of the molecular orbital.

Experimentally, CM  modes are discerned in
nonlinear transport measurements as side-bands in conductance which
correspond to frequencies which do 
not match any of the eigenmodes of a given molecule. 
While the relevance of the separate couplings to two vibrational modes
in model~II is
intuitively clear, the relevance of the simultaneous modulation of
tunneling and coupling to the electron density in model~I is
less clear and deserves further exploration. We  note
that the model~I could be relevant for the nonlinear transport
measurement where the electric 
field would directly couple to the electron density at the molecule,
or in the presence of some impurities with residual electrical fields
even in the (near) linear transport measurements.

The case of a molecule without the CM oscillations described by the
Anderson-Holstein Hamiltonian has already been analyzed by several
authors (see, \emph{e.g.} Ref.~\onlinecite{mravlje05} and references
therein). The dominant effect of the coupling to the
breathing mode is the reduction of the repulsive
interaction among electron pairs occupying the molecular orbital. When
the reduced repulsion $U_\mathrm{eff}$ is positive, a gate-voltage
sweep reveals the area (of width of order $U_\mathrm{eff}$) of enhanced
linear conductance due to the Kondo effect. With increased electron-phonon coupling
and thereby decreased $U_\mathrm{eff}$ this
area is diminished. In this paper we show that the CM
oscillations suppress the conductance in a new way which cannot be
described by an effective model with reduced repulsion. 

This paper is organised as follows. Section II introduces two related
electron-phonon models. In Sec. III numerical methods based on the variational wave function are
presented and some symmetry properties of the models are considered. In Sec. IV results
for linear conductance for both models are presented and some ground state properties of the 
molecular system are explained. Results are summarised in Sec. V.

\section{Models}
Hamiltonians for both models consist of the electron, the phonon and the electron-phonon
coupling terms, $H=H_\mathrm{e}+H_\mathrm{p}+H_\mathrm{e-p}$. In the 
electronic part of the Hamiltonian,
\begin{equation}
H_{\mathrm{e}}=H_\mathrm{L}+H_\mathrm{R}+\epsilon_{d}n_{d}+Un_{d\uparrow}n_{d\downarrow}-t'(\widehat{v}_\mathrm{L}
+ \widehat{v}_\mathrm{R}),
\label{eq:he}\end{equation}
$H_{\mathrm{L,R}}=-t\sum_{i\in\mathrm{L,R}}c_i^\dagger c_{i+1} +
  h.c.$ describe the noninteracting tight-binding left and right leads, respectively, and $\epsilon_d$ is the energy level of the molecular orbital
  occupied by $n=\langle n_d\rangle$ electrons where $n_{d}=\sum_{\sigma}n_{d\sigma}$ with
$n_{d\sigma}=d_{\sigma}^{\dagger}d_{\sigma}$. The characteristic scale
  of the repulsive interaction is $U$ and the tunneling part with 
  characteristic tunneling rate $t'$ 
  consists of operators
  $\widehat{v}_l=\sum_\sigma c_{1l\sigma}^\dagger d_\sigma + h.c.$ for $l=\mathrm{L,R}$ which
  couple the molecular orbital to the first site in the left and right
  leads, respectively. In $H_\mathrm{e}$ only the coupling between the molecular
orbital and even combinations of the left and the right lead orbitals (i.e., the symmetric channel) is present.

Vibrational modes are described with $H^\mathrm{I}_{\mathrm{p}}=\Omega_1
a_1^\dag a_1$ and $H^\mathrm{II}_{\mathrm{p}}=\Omega_1 a_1^\dag a_1 + \Omega_2 a_2^\dag
a_2$ for models I and II, respectively, and  $a_{1,2}^\dagger$ are the
phonon creation operators. The corresponding electron-phonon interaction is given by
\begin{eqnarray} \mathrm{I:}\quad H^\mathrm{I}_{\mathrm{e-p}}=t'g
(\widehat{v}_\mathrm{L}-\widehat{v}_\mathrm{R})x_1 + M (n_d-1)x_1   \:{} \nonumber \\
 \mathrm{II:} \quad H^\mathrm{II}_{\mathrm{e-p}}= t'g
(\widehat{v}_\mathrm{L}-\widehat{v}_\mathrm{R})x_1+M (n_d-1)x_2 , \end{eqnarray} where
$x_{1,2}=a_{1,2}^\dag+a_{1,2}$ are the displacement operators. 
\section{METHODS}

The ground state properties are determined using
the Gunnarsson and Sch\"{o}nhammer
projection-operator method \cite{schonhammer76, gunnarsson85, mravlje05}. Here the Hamiltonian is diagonalized
in the basis
\begin{equation}
|\Psi_{\lambda\left\{ \! m_{\alpha}\!\right\}}\rangle=P_{\lambda}\prod_{\alpha}a_{\alpha}^{\dagger m_{\alpha}}
\left|\tilde{0}\right\rangle ,\label{eq:psi}\end{equation}
which consists of projectors $P_{\lambda}$ , $P_{0}=\left(1-n_{d\uparrow}\right)\left(1-n_{d\downarrow}\right)$,
$P_{1}=\sum_{\sigma}n_{d\sigma}\left(1-n_{d\bar{\sigma}}\right)$,
and $P_{2}=n_{d\uparrow}n_{d\downarrow}$ and additional operators
involving the operators in leads (for example,
$P_3=P_0\widehat{v}_\mathrm{L} P_1$), which are applied to the
state $\left|\tilde{0}\right\rangle$ corresponding to the phonon
vacuum and the ground state of the auxiliary noninteracting Hamiltonian
$\tilde{H}=\tilde{\epsilon}_d n_d +
\tilde{t}'_\mathrm{L} \widehat{v}_\mathrm{L} + \tilde{t}'_\mathrm{R}
\widehat{v}_\mathrm{R} +H_\mathrm{L}+H_\mathrm{R}$ with renormalized local 
energy $\tilde{\epsilon}_d$ and hopping parameters $\tilde{t}'_{\mathrm{L,R}}$.
In the model I  $\alpha=1$ and $\alpha=1,2$ for the model II.  Due to
the CM displacement which induces the coupling to the asymmetric
channel, the renormalized couplings to the left and right lead are
not necessarily 
equal. The chemical potential was set to the middle of the band. A
sufficient number of variational  operators   
(up to $\sim 40$) 
and excited phonon states (up to $\sim 40$) were used in order to
obtain converged results in the parameter regimes presented here.

Zero temperature conductance was calculated by two related methods, both
based on the ground state wave function expressed in the basis Eq.~(\ref{eq:psi}). In the
first method it was calculated from the sine formula (SF) \cite{rejec03b,rejec03a},
$G=G_{0}\sin^{2}[(E_{+}-E_{-})N/4t]$, 
where $G_{0}=2e^{2}/h$ and $E_{\pm}$ are the ground state energies
of a large auxiliary ring consisted of $N$ non-interacting sites
and an embedded interacting system (molecule), with periodic and anti-periodic
boundary conditions, respectively. In the second method we use the
effective parameters $\tilde{t}'_\mathrm{L,R},\tilde{\epsilon}_d$ of the
auxiliary effective noninteracting Hamiltonian $\tilde{H}$ and the conductance
is then given from the corresponding Green's function (GF)
\cite{rejec03b}.  The advantage of the GF method is
that the corresponding  GF can be evaluated for infinite leads so the
difficulties with convergence of SF with number of sites are
avoided.  However, the SF method is robust and it depends only on the
accuracy of the ground-state energy which 
improves with the size of the basis in a transparent way. By comparing
results of both 
methods we checked for the consistency and the convergence. 
From the effective parameters the conductance can  also be calculated
using the Friedel sum rule valid in the limit of large
bandwidth ($t \gg U, M, \Omega_i, ...$). When
the coupling of the system between left and right lead differ,
conveniently parameterized as
$\tilde{t}'_\mathrm{L}=(1+b)\tilde{t}'$, $\tilde{t}'_\mathrm{R}=(1-b)\tilde{t}'$,
the conductance is given by
\begin{equation} \frac{G}{G_0}=\frac{(1-b^2)^2}{(1+b^2)^2}\sin^2{\frac{\pi}{2}}
  n. \label{friedel}\end{equation} 
Note that conductance may be less than unity when both the symmetric
and the antisymmetric channel participate in the transport (\emph{i.e.}, when
$b\neq 0 \land b \neq \pm\infty$). 

The asymmetry stems from the coupling
to the CM displacement, hence the expectation values of
the CM displacement $x=\langle x_1 \rangle$, and the calculated asymmetry factor $b$  
are related. More precisely, the
following exact  \cite{schonhammer84, cornaglia05a} relations hold: \begin{eqnarray} 
\mathrm{I:} \, \langle x_1\rangle=
-\frac{2gt'}{\Omega_1}\langle\widehat{v}_\mathrm{L}-\widehat{v}_\mathrm{R}\rangle-\frac{2M}{\Omega_1}(n-1)
\qquad \qquad  {} \nonumber \\ 
\mathrm{II:} \, \langle x_1
\rangle=-\frac{2gt'}{\Omega_1}\langle\widehat{v}_\mathrm{L}-\widehat{v}_\mathrm{R}\rangle,
\,\, \langle x_2 \rangle=-\frac{2M}{\Omega_2}(n-1) \label{eq:disp}.
\end{eqnarray}
 When
$n=1$ the displacement $x$ is proportional to 
the activity of the antisymmetric channel $\langle
\widehat{v}_\mathrm{L}-\widehat{v}_\mathrm{R}\rangle$ for both models.    

Some general conclusions   may already be
drawn  from considering the particle-hole symmetry. Model I
is invariant with respect to the  
particle-hole transformation
$c_{i\sigma}^\dagger \to c_{i\sigma},d^{\dagger} \to d$ with
$(\epsilon+U/2) \to - (\epsilon+U/2)$, $t\to -t$, $t'\to -t'$ and
$x_1\to-x_1$. Hence at the point of particle-hole
symmetry (where $n=1$), $x=0$, therefore $b=0$ and conductance is
unity according to the Friedel sum rule Eq.~(\ref{friedel}). 
The particle-hole transformation of model II yields $x_2 \to -x_2$
with no bounds on $x_1$. The conductance of model II in the symmetric
point therefore cannot be deduced from considering the particle-hole
symmetry alone and may be less than unity there.

\begin{figure}
\begin{center}\includegraphics[%
  width=70mm,
  keepaspectratio]{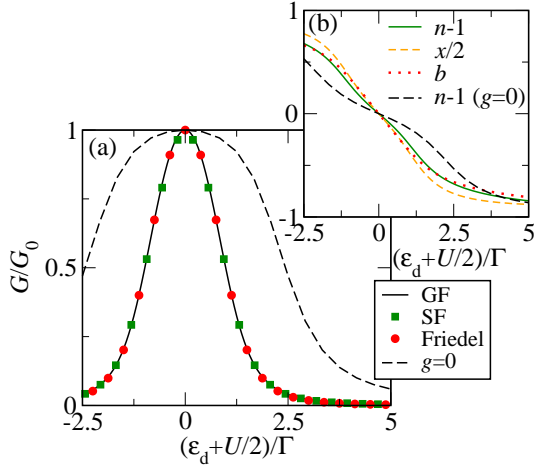}\end{center}

\caption{\label{Fig2} (Color online) (a) Conductance as
  calculated from the Green's function corresponding to 
renormalised parameters (GF, full line),
  sine formula
  method with $N=3000$ (squares), and Friedel sum rule (circles) for molecules described by model I with
  $g=0.4$. For comparison, conductance
  for a molecule without coupling to the CM mode, $g=0$ (dashed).
  (b) Local charge occupancy $n-1$ (full), the displacement
  $x/2$ (short-dashed) the asymmetry parameter
  $b$ (dotted) for a molecule
  described by model I with $g=0.4$, and the occupancy
  $n-1$ for $g=0$ (long-dashed).
}
\end{figure}
 \section{NUMERICAL RESULTS}
Throughout this paper we analyze the system with positive
$U_\mathrm{eff}=U-2M^2/\Omega_\alpha$ and show the results for
$U=10\Gamma$, $M=\Omega_\alpha=2.5\Gamma$, 
$U_\mathrm{eff}=5\Gamma,\Gamma/t=2t'^2/t^2=0.08$, where $\alpha=1,2$ for model I and
II, respectively. The scales of the values used correspond to typical
molecular transistors \cite{yu04_2}. 

 \subsection{Model I}

In Fig.~\ref{Fig2}(a) we show that all
methods yield the same result for conductance within the
model I (the same is true for model II discussed later).  The area of
enhanced conductance is diminished in comparison to the 
$g=0$ case.  In general,
the conductance curves calculated using different methods are 
very close.
However,  due to the finite
bandwidth used here minor discrepancies in results applying the Friedel sum rule, 
Eq.~(\ref{friedel}), might occur when compared to
other methods in the charge transfer 
and empty orbital regimes, $\epsilon_d+U/2\gtrsim t$. 

\begin{figure}
\includegraphics[%
  width=75mm]{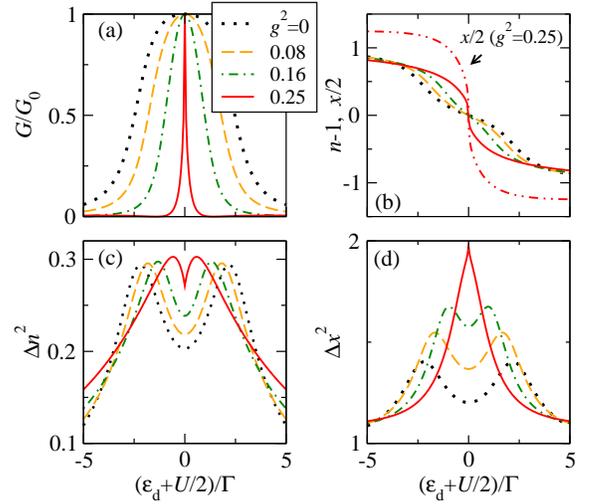}
\caption{(Color online) \label{Fig3}(a) Conductance of a molecule
  described by model I for various CM couplings $g$ and fixed Holstein
  coupling $M$, $2M^2/\Omega_1=U/2$, $U=10\Gamma$, $M=\Omega_1=2.5\Gamma$. (b)
  The occupancy $n-1$ for various $g$. For $g^2=0.25$ also the average displacement $x/2$ (dashed-double-dotted)
  plotted. (c) Fluctuations of charge $\Delta
  n^2=\langle(n_d-n)^2\rangle$ and
 (d) displacement $\Delta x^2=\langle (x_1-x)^2
  \rangle$. 
}
\end{figure}

The results for the asymmetry factor $b$, the 
occupancy $n-1$ and the displacement $x$ are shown
Fig.~\ref{Fig2}(b).  As discussed above, the 
displacement at the symmetric point has to vanish. Correspondingly,
the asymmetry $b=0$ there. Near the
particle-hole symmetric point we find the relation between the
displacement and the asymmetry factor to be approximately linear.

We now systematically explain the properties of the ground state
for model I starting
from a Hamiltonian with $g=0$. In Fig.~\ref{Fig3}(a) we plot the conductance. 
Surprisingly, the influence of increasing $g$ on the conductance is similar to 
simple reduction of $U$ as is the case with the
Anderson-Holstein model where the electron-phonon coupling term $M$ 
effectively reduces $U$.
However, as shown in Fig.~\ref{Fig3}(b) the occupancy $n$ does not exhibit the sharp
transition from $n\sim 2$ to $n\sim 0$ as seen in the Anderson model with
reduced $U_\mathrm{eff}$, where electron (or hole) pairs are preferred in the ground
state (see \emph{e.g.} Ref.~\onlinecite{mravlje05}) and the transport
is dominated by the pair tunneling \cite{koch06}. In the strong coupling
regime the occupancy preferentially takes on values near half integer instead.
 This is because to the lowest order only the average
displacement couples the electronic and 
phononic part, and the average displacement vanishes when $n=1$. Also,
for states with well defined occupancy $n\sim 0,1,2$, the hopping
matrix element has a vanishing weight.

To illustrate this further we determined the charge fluctuations on
the molecule. In general the range corresponding
to increased charge fluctuations $\Delta n^2$ for $g>0$ 
[Fig.~\ref{Fig3}(c)] is extended compared to the 
$g=0$ case. For large $g$ the charge fluctuations still exhibit a
minimum at the
particle hole symmetric point,  in contrast to the negative-$U$ (corresponding to strong
coupling to the breathing mode) case where the charge analog of the
Kondo effect   with
increased charge fluctuations evolves \cite{mravlje05}. The straightforward application
of the fluctuation-dissipation theorem relates $\Delta
n^2\sim -\partial n/ \partial \epsilon_d$, which does not apply to this
system in some parameter regimes.
On the other hand, the displacement fluctuations shown Fig.~\ref{Fig3}(d) are increased
for large $g$ at the symmetric point as in
the Anderson-Holstein model  \cite{mravlje05}.

In Fig.~\ref{Fig3}(b) we also plot
the average displacement $x/2$ of the molecule for $g=0.5$. According to
the relation in Eq.~(\ref{eq:disp}), the difference between the
occupancy (times $2M/\Omega_1$) and
the displacement $x$ is determined by the difference between
expectation values of the tunneling into left and right leads. At the
symmetric point, this difference vanishes.

To summarize, in the particle-hole symmetric point the conductance is
unity, contrary to the corresponding results presented in
Ref.~\onlinecite{alhassanieh05} where a dip in conductance was reported
and related  to the non-applicability of the Friedel sum rule due to
the breakdown of the Fermi liquid. Our results confirm that the conductance can be
expressed with the Friedel sum rule, if left-right asymmetry is
correctly taken into account, Eq.~(\ref{friedel}). The
displacement of the oscillator in model I is determined by the
occupancy. The displacement increases coupling to the anti-symmetric
channel and thereby the charge fluctuations which destroy the Kondo
correlations with well defined charge and enhanced
conductance. Further, the
displacement gives rise to asymmetry, which
means conductance is additionally suppressed for general $\epsilon_d \neq -U/2$.
The width of the conductance peak is seen to decrease approximately as
$U-2M^2/\Omega_1-f(M,\Omega_1,t')g^2$ until this quantity is
reduced to zero. 

\begin{figure}
\includegraphics[%
  width=75mm,
  keepaspectratio]{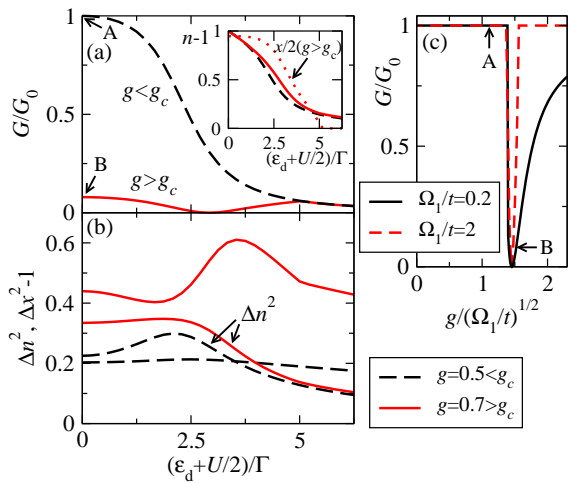}
\caption{(Color online) \label{Fig4}
(a) Conductance of a molecule described by model II for 
  $g<g_c$ (dashed) and $g_c<g<g'_c$ (full). $\Omega_1=0.2t$. Inset: The
  occupancy for the two cases. For $g>g_c$ also the CM displacement
  $x$ is shown (dotted). For $g<g_c$ the CM displacement vanishes, $x=0$. (b)
  Fluctuations of charge and of CM displacement. The curve of charge
  fluctuations for $g=0.5$ is valid for all $g<g_c$. (c) The
  conductance in the symmetric point as a function of $g$. }
\end{figure}

 \subsection{Model II}
We now turn to the results within model II. For small enough values (to be
quantified below) of the CM coupling $g < g_c$ the conductance is not
significantly affected as unlike in model I the 
asymmetry is not determined (directly) by the occupancy of 
the molecule. The average displacement vanishes, correspondingly there
is no asymmetry in the effective model; the width as well as the shape
of the conductance peak and also the correlations
exhibit only minor renormalisation compared to the $g=0$
case. Conversely, for 
$g > g_c$ a soft mode emerges and  
the model has a transition reminiscent of dynamical Jahn-Teller distortion
\cite{balseiro06} as within a simplified treatment of the related model 
 \cite{alascio88}. This is signaled in our approach with non-vanishing
 $x$ and  asymmetry $b$ in the effective model $\tilde{H}$ and
 correspondingly with
 qualitatively modified correlations. For this regime
 additional variational basis functions might be required\cite{hewson80}. For CM coupling
 large enough, $g > g_c'$, only the asymmetric channel is 
active. This is a consequence of the renormalization: 
only the channel with the strongest Kondo coupling remains active at low
temperatures, in accord with Ref.~\onlinecite{balseiro06}, and with corresponding unity conductance.

In Fig.~\ref{Fig4} (a) we plot the conductance and
in Fig.~\ref{Fig4} (b) the fluctuations of occupancy and the
displacement as a
function of $\epsilon_d$ for two distinct $g_c<g<g_c'$ and $g<g_c$ and
fixed $\Omega_1/t=0.2$. Whereas
for $g<g_c$ the conductance and correlations are almost unaffected by
the CM coupling (compare with the $g=0$ curves in Figs.~\ref{Fig2},~\ref{Fig3}),
the $g>g_c$ result is markedly different. In the latter, the
conductance is less than unity at
the symmetric point and is zero at the point corresponding to
$b=1$. In the inset of Fig.~\ref{Fig4} (b) we also plot the occupancy
and for $g>g_c$ the CM displacement. For $\epsilon_d$
large enough, the CM displacement (and correspondingly the asymmetry) is
zero. The precise value for $g_c$ hence varies with the gate voltage
and is related to the charge fluctuations of the bare model: increased
charge fluctuations induce a transition to the state of 
spontaneously broken symmetry.

In Fig.~\ref{Fig4} (c) we plot the conductance as a function of
coupling to the CM mode. For small $g$ the
conductance is unity as only the even channel is active, $b=0$. By
increasing $g$, the conductance drops sharply at $g_c$ which is found to be set by
the relation $g_c^2 
t/\Omega_1=c(t'),$  where $c(t')\sim 2.5$ for the results presented
here \cite{rel}. Note labels A and B indicating the values of $g$
for which the results are shown in
Figs.~\ref{Fig4}(a,b). For large $\Omega_1$ both channels are active with corresponding
suppressed conductance only for a narrow range of $g>g_c$  while for
progressively decreasing $\Omega_1$ the range of $g$ with suppressed
conductance widens. As in typical devices considered in
experiments\cite{yu04_2} $\Omega_1<\Gamma<U$ the anomalies in conductance
of the type considered here would would not be sensitive to
the precise value of the coupling to the CM mode provided that it is
large enough.

\section{CONCLUSIONS}
In conclusion, we analyzed the properties of molecules with
a center-of-mass vibrational mode in linear transport measurements
described by two distinct Hamiltonians. In the first case, where the
coupling to the CM mode 
simultaneously modulates the molecular energy level, the result of the
coupling to the CM mode is a reduced range of gate voltages
with enhanced conductance. As a direct consequence of the
particle-hole symmetry the conductance in the symmetric point remains unity.

In the second case where the coupling to the CM mode modulates only the
tunneling matrix elements and additional breathing mode couples to the
electron density, the CM mode is of minor relevance for the
transport  until a critical value of
this coupling $g_c$ is reached. For values of the coupling near 
the critical value anomalies due to the development of the soft mode arise
signaled by the conductance which is less than unity.
The conductance  for these values is within the effective model suppressed for
a wide range of gate voltages.

We explained the relation of CM displacement to 
the transport of electrons for both models and substantiated the explanation by the
numerical data for occupancy, occupancy fluctuations and also related
phonon expectation values.

Finally, we may make some remarks regarding the experimental
relevance of results presented here. 
For linear regime without residual electric fields it does
not seem likely that the model I would provide an accurate description
of the experimental situation. By analyzing model II which applies
then, we have shown that for small couplings to the CM mode there are
no discernible effects on the linear conductance (we should note here
that this statement would not be changed if additional breathing modes
were introduced). At a sufficient coupling strength this picture
changes: the displacement of the system from the point of inversion
symmetry becomes increasingly sensitive to external perturbations
which would result in a displacement of molecule and strong
suppression of conductance.

To check for this behaviour in experiments would require the following
steps. First devices in which CM side-peaks are seen would have to be
produced, possibly by using different molecules, which would give in
other respects identical devices besides having different frequencies
of the CM motion. The fingerprint of the physics considered here would
then be the suppression of conductance which would predominantly occur
for the devices with lower frequencies of the CM motion.  As this
procedure would require near-atomic similar lead configuration near
the molecule the systematic experimental confirmation of the analyzed
effects seems currently hard to achieve. We speculate nevertheless
that this effects might have contributed to the anomalously suppressed
conductance in some of the devices considered in, \emph{e.g.},
Ref.~\onlinecite{pasupathy05}.

\acknowledgements{
 We thank R. \v{Z}itko for inspiring suggestions and S.~El~Shawish for
 discussions regarding the Jahn-Teller effect. The work was supported
 by SRA under grant Pl-0044.   
}

\end{document}